\documentclass[11pt,a4paper,twocolumn]{article}
\usepackage{jssij}

\title{Development of a Tracklet Extraction Engine}
\author[1]{Ryou Ohsawa$^{*1}$}

\begin{document}
\maketitle

\begin{abstract}
An efficient algorithm is required to extract moving objects (asteroids, satellites, and space debris) from enormous data with advances in observational instruments. We have developed an algorithm, \texttt{tracee}, to swiftly detect points aligned as a line segment from a three-dimensional space. The algorithm consists of two steps; First, construct a $k$-nearest neighbor graph of given points, and then extract colinear line segments by grouping. The proposed algorithm is robust against distractors and works properly even when line segments are crossed. While the algorithm is originally developed for moving object detection, it can be used for other purposes.
\end{abstract}


\section{Introduction}
\label{sec:intro}

Solar system small bodies are relatively small objects orbiting around the Sun, such as asteroids, comets, and meteoroids. More than one million objects have been discovered as of August 2020\footnote{The number is adopted from the Minor Planet Center (\today).}. The objects whose perihelion distances are smaller than 1.3 au are called Near-Earth Objects (NEOs). Their population, orbits, sizes, and compositions provide vital information to understanding the solar system's evolution \citep[e.g.,][]{granvik_debiased_2018,bottke_debiased_2002}. Besides, some of them are potentially hazardous to the Earth. Thus, detecting NEOs is crucial as well in the context of planetary defense. However, many NEOs, especially smaller ones, remain undiscovered \citep[e.g.,][]{harris_population_2015}.

Multiple images of the same field are obtained with some intervals to detect moving objects in optical observations. Moving objects should appear in different positions of the images. A short arc is obtained by connecting such two sources, usually referred to as a tracklet. Connecting multiple tracklets provides a longer track, leading to a more precise orbit determination of a moving object.

Algorithms connecting tracklets have been developed so far. Moving Object Processing System \citep[MOPS;][]{denneau_pan-starrs_2013,kubica_efficient_2007} is the algorithm used in Pan-STARRS, utilizing a tree-based structure to efficiently connect inter-night tracklets. ZTF's Moving Object Discovery Engine \citep[ZMODE;][]{masci_zwicky_2019} is another implementation for Zwicky Transient Facility. ZMODE constructs a "stringlet" \citep{waszczak_main-belt_2013} from three or more detections, making the system more robust. HelioLinc provides a new idea to this field, where tracklets are connected in a heliocentric frame instead of a geocentric one. Although some algorithms are independent of tracklets \citep[THOR;][]{moeyens_thor_2021}, extracting tracklets is the initial step in moving object discovery.

The cadence of survey observations has increased with advances in observational instruments \citep[e.g., Pan-STARRS, ZFS, ATLAS, ASAS-SN, and EvryScope;][]{chambers_pan-starrs1_2016,magnier_pan-starrs_2016,dekany_zwicky_2020,tonry_atlas:_2018,kochanek_all-sky_2017,law_evryscope_2015}. Video observation is an extreme case \citep[e.g., Tomo-e Gozen, OASES, and TAOS2;][]{sako_tomo-e_2018,arimatsu_organized_2017,huang_taos_2021,lehner_status_2018}. An efficient data reduction process is required due to the enormous amount of data in the video observation. Parallel computing is one solution. \citet{yanagisawa_automatic_2005} developed an FPGA board to extract faint tracklets from a sequence of images. This paper presents another solution, developing an efficient algorithm. The proposed algorithm extracts linearly aligned points in a three-dimensional space, a fundamental routine to extract tracklets from video data.

The paper is organized as follows. The design and requirements of the proposed algorithm are described in Section 2. Then, the performances of the algorithm are illustrated in Section 3. Section 4 summarizes the paper. The algorithm we develop is available as open-source software.

\section{Algorithm}
\label{sec:algorithm}

This study aims to accelerate a process to identify moving object candidates from a sequence of images. Many sophisticated applications are available to extract sources from images \citep[e.g., \texttt{DAOFIND} and \texttt{Source Extractor}; ][]{stetson_daophot_1987,bertin_sextractor:_1996}. The list of the extracted sources may contain fixed stars, cosmic rays, and artificial noises, as well as possible moving objects. It is not an easy task to isolate moving object candidates from such distractors. This process can be a bottleneck in moving object extraction.

We developed a dedicated algorithm, Tracklet Extraction Engine (\texttt{tracee}), which efficiently extracts moving object candidates from a source list. The \texttt{tracee} is originally developed for an asteroid search but is possibly utilized in other fields. Table~\ref{tab:appendix:tutorial} describes a basic usage of \texttt{tracee}.

\subsection{Problem Setting}
\label{sec:algorithm:problem}

The algorithm we develop, \texttt{tracee},  is intended to deal with short video data. Thus, we assume that the motions of moving objects are linear and uniform. Thus, the problem is described as extracting linearly aligned points from a point cloud in a three-dimensional space. A single data set may contain multiple moving objects. The algorithm is required to detect them separately. The moving objects possibly cross fixed stars and each other. The algorithm should work properly in such cases.

\textit{Hough transformation} \citep{hough_method_1962} a plausible algorithm to deal with the problem, where each source votes for points in the hyperspace where each point is associated with a line segment. Such voting-based algorithms are, however, usually memory-consuming, while some efficient implementations are proposed \citep{li_fast_1986}. Thus, we do not adopt Hough transformation. Instead, we split the problem into two parts; First, we construct a $k$-nearest neighbor graph of the sources and then extract linearly aligned edges from the graph. An overview of the procedures is presented in Algorithm~\ref{algorithm:tracee}. Detailed descriptions are described below.

\begin{algorithm}[tp]
\caption{Overview of \texttt{tracee}}
\label{algorithm:tracee}
\DontPrintSemicolon
\SetKwProg{Part}{}{}{}
\Part{\textbf{\upshape Part 1:} $k$-Nearest Neighbor Graph}{
\Input{$V=\left\{v_i\middle|_{i=1{\ldots}N}\right\}$}
\Output{$E=\left\{(v_i,v_j)\middle|_{i,j \in \{1{\ldots}N\}}\right\}$}
{Construct a $k$-NN graph $G(V,\,E)$ from $V$.}\;
{Merge the bidirectional edges in $E$.}\;
{Redirect the edges in the chronological order.}\;
{Remove inappropriate edges in $E$.}\;
\KwRet{$E$}\;
}
\BlankLine
\Part{\textbf{\upshape Part 2:} Line Segment Grouping}{
\Input{$E=\left\{(v_i,v_j)\middle|_{i,j \in \{1{\ldots}N\}}\right\}$}
\Output{$B=\left\{b_m\middle|_{m=1{\ldots}M}\right\}$}
{Construct a set of baselines $B$ from $E$.}\;
{Remove short baselines from $B$.}\;
{Remove baselines with large scatter from $B$.}
\KwRet{$B$}\;
}
\end{algorithm}

\subsection{Procedures}
\label{sec:algorithm:procedure}

\subsubsection{Input and Output}
\label{sec:algorithm:procedure:input/output}

The \texttt{tracee} receives a list containing the source positions and the timestamps of extraction. Hereafter, we call an element of the list a \textit{vertex}. The input data set $V$ is described as follows:
\begin{equation}
\label{eq:algo:vertex}
V = \Bigl\{
v_i = \left\{ x_i, \, y_i, \, t_i \right\} \Bigm|
i = 0,\ldots,N
\Bigr\},
\end{equation}
where $i$ is the index of the source, $v_i$ is the vertex corresponding to the $i$th source, $x_i$ and $y_i$ are the source positions, $t_i$ is the extraction timestamp, and $N$ is the total number of the extracted sources.

The tracklet is defined as a line segment or an edge. We define a \textit{baseline} as a data structure containing a line segment and associated vertices:
\begin{equation}
\label{eq:algo:tracklet}
b = \left\{ v^s, v^e, w \right\},
\end{equation}
where $v^s$ and $v^e$ are respectively the starting and end points of the line segment, $w$ contains the list of the vertices associated with the baseline. The definitions of these elements are described below:
\begin{equation}
\label{eq:algo:tracklet:detail}
\begin{cases}
v_s = \left\{ x^s,\, y^s,\, t^s \right\}, \\
v_e = \left\{ x^e,\, y^e,\, t^e \right\}, \\
w = \left\{ \theta_1,\, \theta_2,\, \ldots,\, \theta_{K} \right\},
\end{cases}
\end{equation}
where $K$ is the number of the vertices associated with the baseline and $\theta_i$ denotes the index of the vertex. The algorithm is expected to return a set of baselines, $B$:
\begin{equation}
\label{eq:algo:baseline}
B = \Bigl\{
b_m = \left\{ v^\mathrm{s}_m,\, v^\mathrm{e}_m,\, w_m \right\}
\Bigm| m = 0,\ldots,M
\Bigr\},
\end{equation}
where $m$ is the index of the baseline and $M$ is the total number of the baselines.

\subsubsection{$k$-Nearest Neighbor Graph}
\label{sec:algorithm:procedure:knn}

\begin{figure}[t]
\centering
\includegraphics[width=\linewidth]{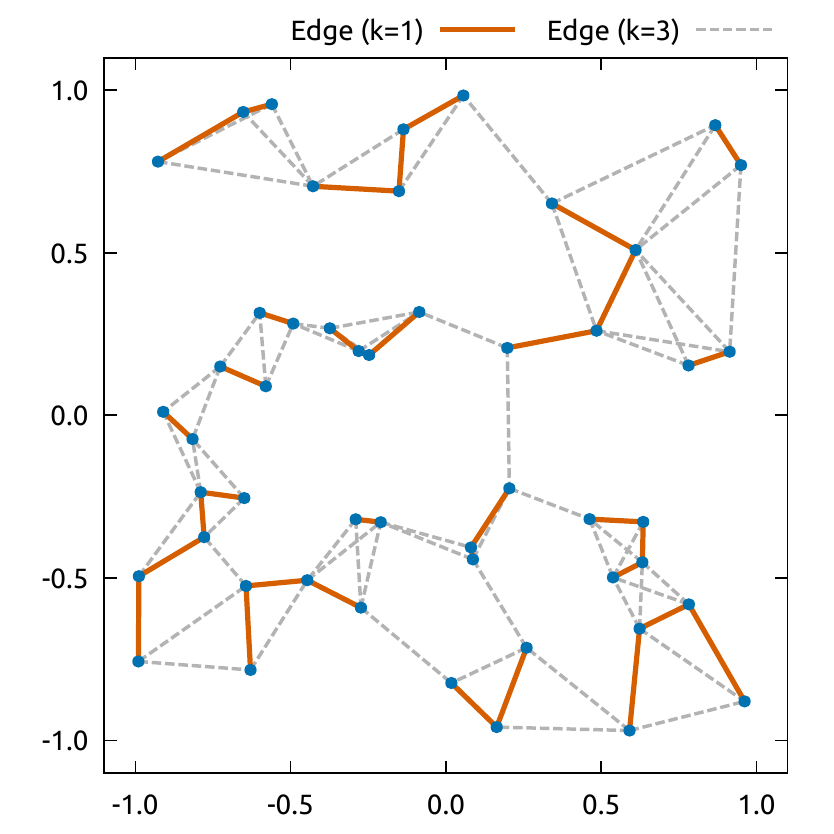}
\caption{A sample of $k$-nearest neighbor graph. The blue dots indicate the vertices. The nearest neighbor graph ($k{=}1$) is denoted by the red thick lines. The $3$-nearest neighbor graph is shown by the gray dashed lines.}
\label{fig:algo:knn_sample}
\end{figure}

In the first part of the algorithm, we construct a $k$-nearest neighbor graph ($k$-NN graph) from the vertices. Every vertex of the $k$-NN graph has at least $k$ edges to the $k$th nearest vertices. Samples of the $k$-NN graph are presented in Figure~\ref{fig:algo:knn_sample}. The blue dots represent the vertices. The solid and dashed lines show the $k$-NN graphs for $k=1$ and $3$, respectively.

A brute-force construction of a $k$-NN graph requires a cost of $O(n^{2})$, which is not acceptable for large problems. Dong et al. (2011) presented \texttt{NN-Descent}, an efficient algorithm to obtain an approximate $k$-NN graph, where the construction cost is about $O(n^{1.14})$ and is suitable for parallel computing. An overview of \texttt{NN-Descent} is described in Algorithm~\ref{algorithm:nndescent}. The algorithm receives the vertex set $V$ and the order of the graph $k$. The edge list $E[v]$ handles the edges starting from $v$. The element of $E[v]$ is a pair of the destination $w$ and the distance between $v$ and $w$. At first, $E[v]$ is initialized with arbitrary $k$ vertices with infinity distances. Then, the edge list $E$ is iteratively updated by replacing elements with shorter ones. The algorithm stops when no element is updated.

\begin{algorithm}[t]
\caption{\texttt{NN-Descent}}
\label{algorithm:nndescent}
\SetKwFor{Loop}{loop}{}{}
\DontPrintSemicolon
\Input{vertex set $V$, order $k$}
\Output{$k$-NN edge list $E$}
\BlankLine
\ForEach(\tcp*[f]{initialize $E$}){$v \in V$}{
{$W \gets$ arbitrary $K$ elements from $V$.}\;
\lForEach{$w \in W$}{append $\langle w, \infty\rangle$ to $E[v]$.}
}
\Loop(\tcp*[f]{iteratively update $E$}){}{
{$R \gets$ reverse graph of $E$.}\;
\lForEach{$v \in V$}{$\overline{E}[v] \gets E[v] \cup R[v]$.}
{$c \gets \mathit{False}$}\;
\ForEach{$v \in V$}{
\ForEach{$u_1 \in \overline{E}[v]$, $u_2 \in \overline{E}[u_1]$}{
{$l \gets$ distance between $v$ and $u_2$}\;
{$\langle \tilde{v},s \rangle \gets$ largest element of $E[v]$.}\;
\If{$l < s$}{
{pop $\langle \tilde{v},s \rangle$ from $E[v]$.}\;
{append $\langle u_2,l \rangle$ into $E[v]$.}\;
{$c \gets \mathit{True}$}\;
}
}
}
\lIf{$c = \mathit{False}$}{\KwRet{$E$}}
}
\end{algorithm}

\texttt{NN-Descent} has been widely used, and there are already several implementations. While major implementations are well-established and functionally rich, high performance is required by tracee. Thus, we have developed a minimal package to construct a $k$-NN graph efficiently in a three-dimensional Euclidian space \citep{ohsawa_minimalknn_2020}. The core function of the module is written in \texttt{C++}, and an interface for \texttt{Python} is provided. The package, \texttt{minimalKNN}, is available in the Python Package Index as open-source software.

The output of Algorithm~\ref{algorithm:nndescent} is a directed graph. The edges should be in chronological order to trace moving objects. Thus, the bidirectional edges are merged, and the edges are redirected in chronological order. We also remove inappropriate edges whose corresponding velocities are too high. Then, the updated $k$-NN graph is passed to further analysis.

\subsubsection{Line Segment Grouping}
\label{sec:algorithm:procedure:grouping}

Linearly aligned edges of the graph are extracted to identify moving object candidates. \citet{jang_fast_2002} tackled the problem of extracting favorable line segments from a gray-scale image. The proposed algorithm is a fast line segment grouping method; First, edges are extracted by an edge detector. Then, the edges are split into short line segments (hereafter, elementary line segments, ELSs). The algorithm they developed extracts a subset of linearly aligned line segments from a set of ELSs. An overview of the algorithm is described in Algorithm~\ref{algorithm:fdlsgm}.

\begin{algorithm}[t]
\caption{\texttt{fast ELS grouping}}
\label{algorithm:fdlsgm}
\SetKwFor{Loop}{loop}{}{}
\DontPrintSemicolon
\Input{set of ELSs $E$}
\Output{set of baselines $B$}
\BlankLine
{construct an empty \textit{accumulator} $A$.}\;
\ForEach(\tcp*[f]{initialization}){$e \in E$}{
{$\theta \gets \mathrm{argument}(e)$}\;
{$c \gets \mathit{False}$}\;
\ForEach{$b \in A[\theta]$}{
\If{$e \sim b$} {
{append $e$ to the members of $b$}\;
{$c \gets \mathrm{True}$}
}
}
\If{$c = \mathit{False}$}{
{generate a baseline $b'$ from $e$}\;
{append $b'$ to $A[\theta]$.}}
}
\Loop(\tcp*[f]{optimization}){}{
\ForEach{$e \in E$}{
{$\theta \gets \mathrm{argument}(e)$}\;
{$c \gets \mathit{False}$}\;
\ForEach{$b \in A[\theta]$}{
\If{$e \sim b$ \& $e \not\in b$}{
{append $e$ to the members of $b$}\;
{$c \gets \mathrm{True}$}
}
}
}
}
{construct an empty \textit{baseline set} $\overline{B}$}\;
\ForEach(\tcp*[f]{merge baselines}){$b \in B$}{
{$\theta \gets \mathrm{argument}(b)$}\;
\ForEach{$\tilde{b} \in A[\theta]$}{
\If{$b \sim \tilde{b}$}{
{$b \gets \mathrm{merge}(b,\tilde{b})$}\;
{remove $b$ and $\tilde{b}$ from $B$}\;
}
}
{append $b$ into $\overline{B}$}\;
}
{$B \gets \overline{B}$}\;
\end{algorithm}

\begin{figure}
\centering
\includegraphics[width=.9\linewidth]{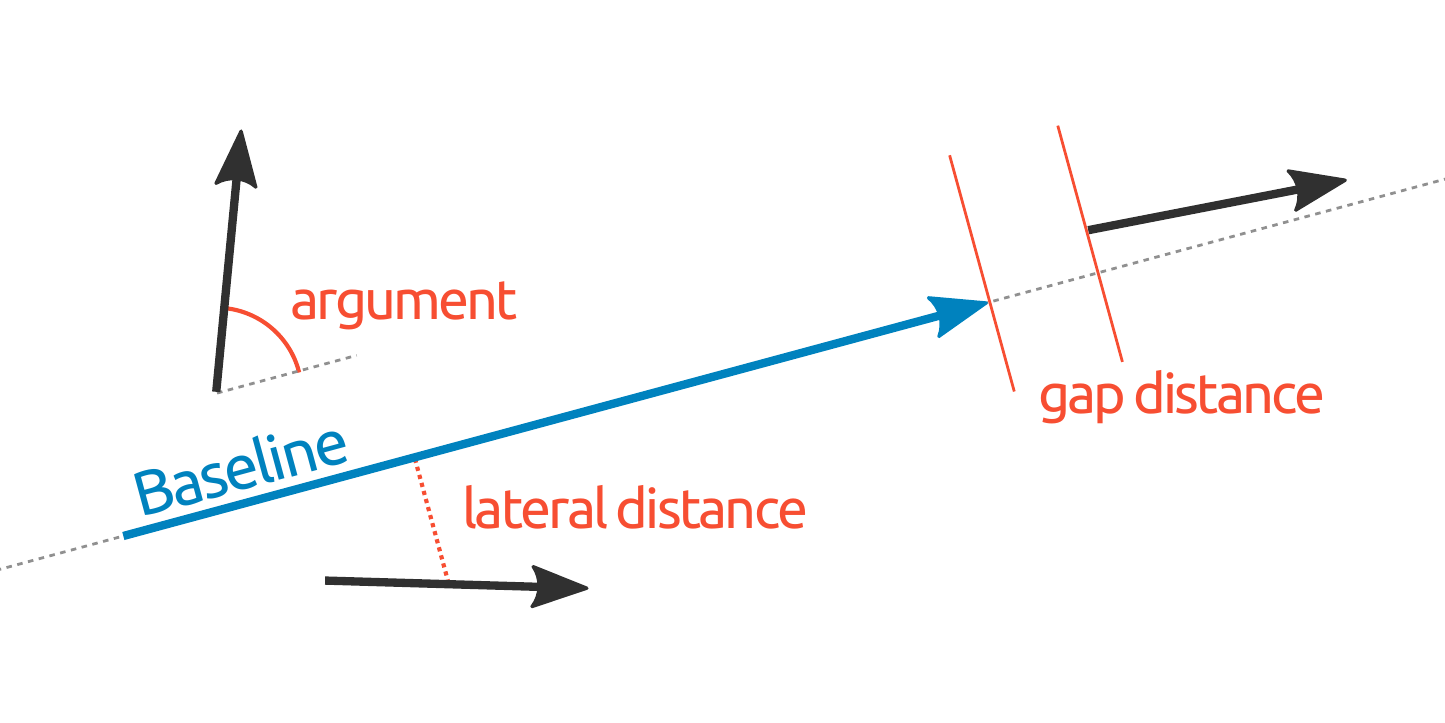}
\caption{A schematic view of the line segment grouping criteria. The blue arrow indicates the baseline, while the short black arrows indicate the elemental line segments.}
\label{fig:algo:fdlsgm:criteria}
\end{figure}

\textit{Baselines} and an \textit{accumulator} play an essential role in facilitating line segment grouping. The Baselines are line segment candidates obtained by grouping colinear ELSs, while the accumulator is a data structure to search baselines by the position angle. First, each ELS is compared with baselines in the accumulator. Then, if the ELS and the baseline fulfill the similarity criterion, the baseline is updated by embracing the ELS. Otherwise, a new baseline is created from the ELS. Refer to \citet{jang_fast_2002} for the way to create and update baselines. The definition of the similarity is illustrated in Figure~\ref{fig:algo:fdlsgm:criteria}. The \textit{argument} is the angle between the ELS and the baseline, the \textit{lateral distance} is the distance perpendicular to the baseline, and the \textit{gap distance} is the separation between the ELS and the baseline projected on the baseline. Second, the same procedure is repeated one more time, but no new baseline is created this time. Finally, similar baselines are merged into a single baseline.

The original algorithm in \citet{jang_fast_2002} works on non-directional line segments on the two-dimensional plane. Thus, we have extended the algorithm to directional line segments in the three-dimensional space. The extended algorithm is named \textit{Fast Directed Line Segment Grouping Method} (\texttt{}fdlsgm). The developed code is published as open-source software \citep{ohsawa_fdlsgm_2020}. The core function of the module is written in C++, and an interface for Python is implemented.

The output of fdlsgm may contain unsuccessful baselines which are stochastically grouped as lines. First, the baselines are removed when the number of associated vertices is small. Then, the scatter of the vertices around the baseline is evaluated, and the baselines with large scatter are removed. The remaining baselines are finally returned as the candidates of tracklets.

\section{Performance Verification}
\label{sec:performance}

The performance of \texttt{tracee} is evaluated using mock data. In Section~\ref{sec:performance:mockdata}, we demonstrate the way that the algorithm works. Tracklets are extracted in several different conditions. The scalability of the algorithm is evaluated in Section~\ref{sec:performance:scalability} with the different numbers of moving objects.

\subsection{Validation with Mock data}
\label{sec:performance:mockdata}

Mock data are generated as follows; First, $n$ artificial moving objects are generated in a field of $512{\times}512\,\mathrm{pixel}$. The velocities and directions of the moving objects are randomly selected. Next, the positions of the moving objects are measured for continuous 30 frames, while they are disturbed assuming the uncertainty of the position measurement of $\sigma = 0.2\,\mathrm{pixel}$. Then, the measurements are stochastically dropped with a probability of $p$. Finally, $m$ distractors are added to the measurements.

The generated mock data are consistently processed with the same parameters. In constructing the $k$-NN graph, the number of neighbors $k$ is set to 10, but only the three shortest edges are adopted. Then, the edges faster than $200\,\mathrm{pixel/frame}$ are removed. In the line segment grouping, the argument threshold is set to $3^\circ$. The lateral distance threshold is set to $1\,\mathrm{pixel}$. The gaps between baselines are accepted up to three times the baseline lengths. Extracted tracklets are removed when they have less than 10 sources.

\begin{figure*}
\centering
\includegraphics[width=.98\linewidth,page=1]{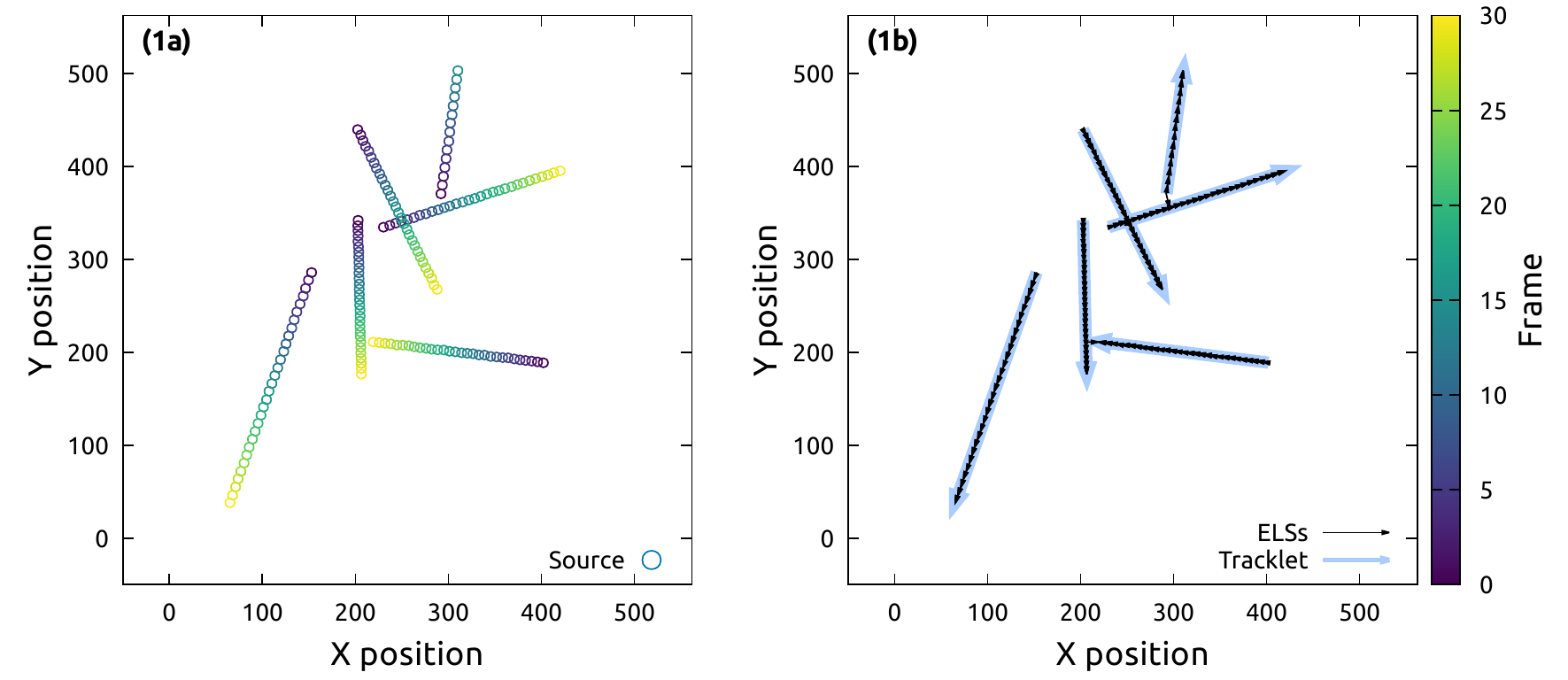}
\includegraphics[width=.98\linewidth,page=2]{f004.pdf}
\includegraphics[width=.98\linewidth,page=3]{f005.pdf}
\caption{Mock data with different situations. The top panels show the data for $(n,p,m)=(6,0,0)$. Panel (1a) shows the source positions projected on the $XY$-plane, where the symbol colors indicate the frame number. Panel (1b) illustrates the generated elemental line segments (ELSs, the thin black arrows) and the extracted tracklets (the thick blue arrows). The middle and bottom panels illustrate the data generated with $(n,p,m)=(6,0.5,0)$ and $(6,0,500)$, respectively.}
\label{fig:performance:mockdata}
\end{figure*}

\begin{figure*}
\centering
\includegraphics[width=.48\linewidth,page=1]{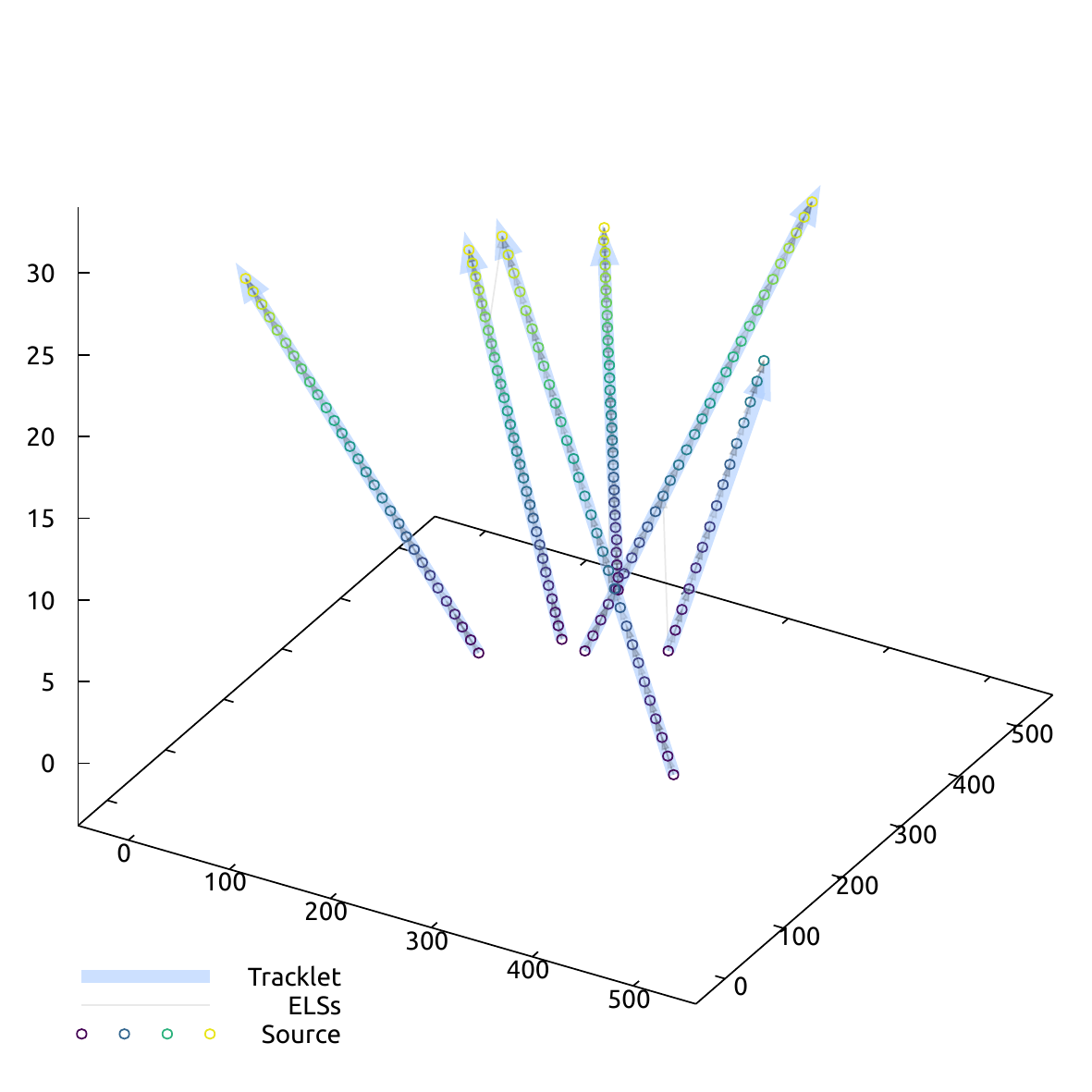}
\includegraphics[width=.48\linewidth,page=2]{f007.pdf}
\includegraphics[width=.48\linewidth,page=3]{f008.pdf}
\caption{A 3-dimensional views of Case 1 (top left), 2 (top right), and 3 (bottom). The source positions are shown by the circle symbols. The thin gray arrows indicates the ELSs. The retrieved tracklets are shown by the thick blue arrows.}
\label{fig:performance:3d}
\end{figure*}

\begin{table*}
\centering
\caption{Summary of Scalability Test}
\label{tab:performance:elapsed}
\begin{tabular}{ccr@{$\pm$}lr@{$\pm$}lr@{$\pm$}l}
\toprule
$N_\mathrm{obj}$
& \multicolumn{1}{c}{$\langle N_\mathrm{track} \rangle$}
& \multicolumn{2}{c}{$\langle N_\mathrm{ELS} \rangle$}
& \multicolumn{2}{c}{$\langle T_\mathrm{ELS} \rangle$}
& \multicolumn{2}{c}{$\langle T_\mathrm{total} \rangle$} \\
\midrule
$5$   & $5$     & $617$  & $17$
& $16.6$ & $0.7\,\mathrm{ms}$
& $32.1$ & $2.2\,\mathrm{ms}$ \\
$10$  & $10$    & $878$  & $40$
& $26.7$ & $2.4\,\mathrm{ms}$
& $60.3$ & $6.6\,\mathrm{ms}$ \\
$25$  & $24.8$  & $1577$ & $37$
& $59.7$ & $2.9\,\mathrm{ms}$
& $134$  & $5.2\,\mathrm{ms}$ \\
$50$  & $49$    & $2830$ & $121$
& $145$  & $7.7\,\mathrm{ms}$
& $356$  & $31\,\mathrm{ms}$ \\
$100$ & $98.2$  & $5272$ & $78$
& $391$  & $27\,\mathrm{ms}$
& $1.05$ & $0.05\,\mathrm{s}$ \\
$125$ & $117.6$ & $6505$ & $167$
& $539$  & $28\,\mathrm{ms}$
& $1.56$ & $0.09\,\mathrm{s}$ \\
$150$ & $136.6$ & $7611$ & $53$
& $755$  & $25\,\mathrm{ms}$
& $2.13$ & $0.10\,\mathrm{s}$ \\
\bottomrule
\end{tabular}
\end{table*}

Mock data of $n = 6$ are illustrated in Figure~\ref{fig:performance:mockdata}. Panels (1a) and (1b) illustrate the result of Case~1, generated with $(n,p,m)=(6,0,0)$. The source distribution projected on the $XY$-plane is shown in Panel~(1b). The ELSs generated from the sources are illustrated by the thin black arrows in Panel~(1b), while the extracted tracklets are shown by the thick blue arrows. All the six moving objects are successfully identified by \texttt{tracee}. A 3-dimensional view of Case 1 is presented in Figure~\ref{fig:performance:3d}.

Panels (2a) and (2b) show the result of Case~2, generated with $(n,p,m) = (6,0.5,0)$. The input moving objects are the same as in the previous case, but about half of the data are randomly removed (see, Panel~(2a)). Although the number of ELSs decreases accordingly, all the tracklets are successfully recovered. A 3-dimensional view of Case 2 is presented in Figure~\ref{fig:performance:3d}.

The result of Case~3, generated with $(n,p,m) = (6,0,500)$, is presented in Panels (3a) and (3b). The moving objects are the same as in Case~1, while 500 distracting vertices are randomly appended. The total number of vertices is 680. Thus, a brute force approach will produce 230860 possible line segments. On the other hand, the number of ELSs in Panel~(3b) is 1217. The algorithm reduces the number of segments to just 0.5\% of those in the brute force approach. The thick blue arrows indicate that all the moving objects are successfully identified as well. A 3-dimensional view of Case 3 is presented in Figure~\ref{fig:performance:3d}.

\subsection{Scalability}
\label{sec:performance:scalability}

The performance of tracee with different numbers of moving objects is investigated. As in Section~\ref{sec:performance:mockdata}, mock data are generated with $n = 5$, $10$, $25$, $50$, $100$, $125$, and $ 150$, while $p$ and $m$ are fixed to 0 and 200, respectively. For each n, five datasets are drawn with different random seeds. All the datasets are processed with the same parameters as in Section~\ref{sec:performance:scalability}. The calculation was conducted on a laptop with Intel Core i7-6600U ($2.60\,\mathrm{GHz}$). The total elapsed time and the elapsed time for creating the k-NN graph are measured using the \texttt{\%timeit} command provided by \texttt{IPython} \citep{perez_ipython_2007}.

The results are summarized in Table~\ref{tab:performance:elapsed}, where $N_\mathrm{obj}$ is the number of moving objects, $\langle N_\mathrm{track} \rangle$ is the mean number of extracted tracklets, $\langle N_\mathrm{ELS} \rangle$ is the mean number of ELSs, $\langle T_\mathrm{ELS} \rangle$ is the mean elapsed time for creating the $k$-NN graph, and $\langle T_\mathrm{total} \rangle$ is the mean total elapsed time. In general, the moving objects are successfully identified in every case. Some objects are missed since they are generated around edges and shortly move outside. The detection rate slightly decreases for larger $N_\mathrm{obj}$, mainly because similar tracklets are wrongly merged. Such tracklets can be distinguished by tuning the threshold parameters, but this is beyond the scope of this paper.

\begin{figure}
\centering
\includegraphics[width=\linewidth]{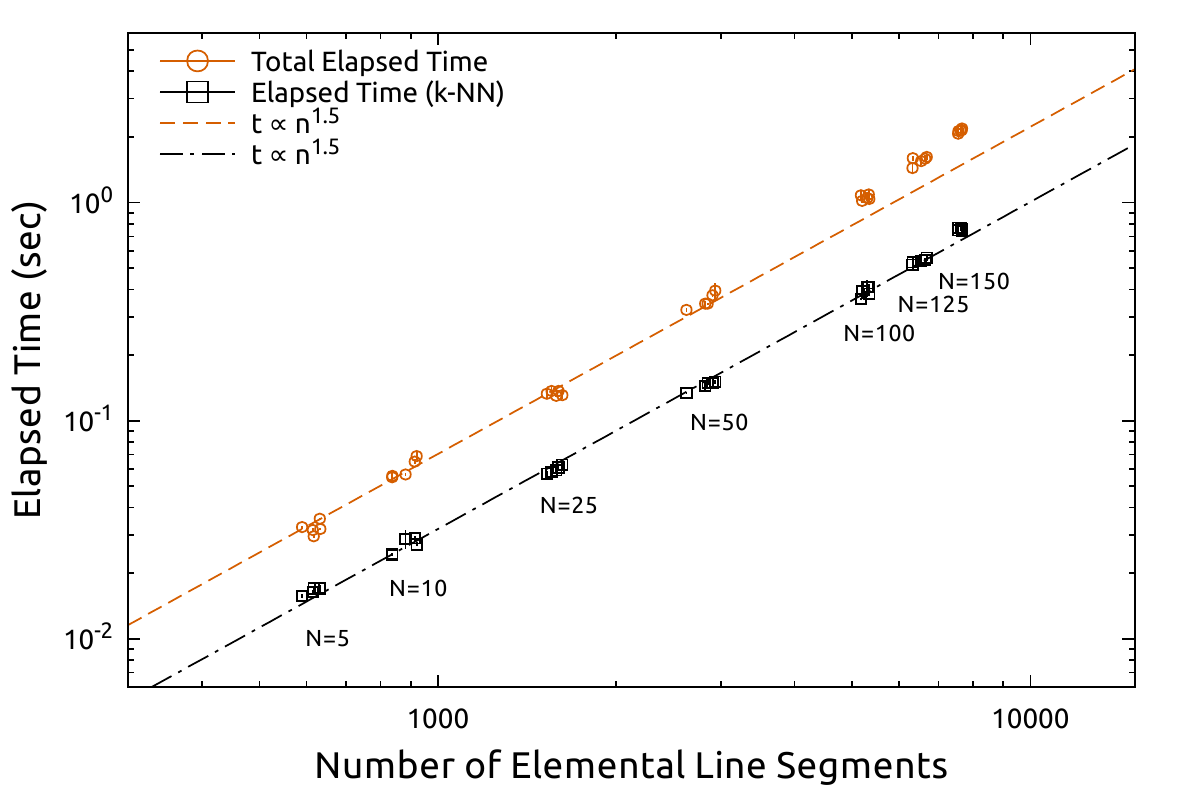}
\caption{Dependency of the elapsed time on the number of moving objects. The red circles show the total elapsed times, while the black squares show the elapsed times for creating the $k$-NN graph. The dashed and dot-dashed lines indicate that $t \propto N_\mathrm{ELS}^{1.5}$.}
\label{fig:performance:scale}
\end{figure}

Figure~\ref{fig:performance:scale} shows the elapsed times against the number of the elemental line segments, which is the most appropriate indicator of the problem size. The red circles show the total elapsed times, while the black squares show the elapsed times for generating the $k$-NN graph. Both elapsed times roughly follow $t \propto N_\mathrm{ELS}^{1.5}$, as shown by the dashed and dot-dashed lines. The figures below the symbols denote the numbers of moving objects. Since $N_\mathrm{ELS}$ is almost proportional to $N_\mathrm{obj}$, the elapsed times approximately follow $t \propto N_\mathrm{obj}$. The same holds for the number of distractors. Thus, Figure~\ref{fig:performance:scale} indicates that the elapsed times increase at most with $N^{1.5}$, where $N$ is the representative size of a problem. There is, however, an excess in the total elapsed time around $N_\mathrm{ELS} \sim 8000$. The excess is possibly attributed to the \textit{accumulator}. The performance of the \textit{accumulator} may decrease for such a large $N_\mathrm{ELS}$. More efficient implementation of the \textit{accumulator} will improve the performance of \texttt{tracee} for larger problems

\subsection{Application to Real Data}
\label{sec:performance:application}

Finally, we cite an application of tracee in the Tomo-e Gozen transient survey to attest that the algorithm can deal with actual observational data. Tomo-e Gozen is a wide-field video camera developed in Kiso Observatory, the University of Tokyo, capable of monitoring a sky of 20 square degrees at up to 2\,fps \citep{sako_tomo-e_2018}. Kiso Observatory conducts a transient survey using Tomo-e Gozen. The survey is intended to detect short transient objects in the entire observable sky. Tomo-e Gozen obtains 6- or 9-second video data in every pointing with changing observation fields to sweep the entire sky.

We have developed a data reduction pipeline to extract moving objects in the video data obtained by Tomo-e Gozen. The \texttt{tracee}'s parameters are tuned to detect near-earth objects with apparent speeds of $v \sim 1''\mathrm{s^{-1}}$. The obtained data are automatically processed by dedicated computers installed on Kiso Observatory within a night. The reduction pipeline successfully detects thousands of moving objects a night, including asteroids, artificial satellites, and space debris. From March 2019 to May 2021,  28 near-earth asteroids have been discovered in the survey, showing the applicability of \texttt{tracee} for actual observational data. Detailed analysis of this pipeline's performance will be presented in a forthcoming paper (Ohsawa et al., \textit{in prep}).

\section{Conclusion}
\label{sec:conclusion}

We have developed an efficient algorithm, \texttt{tracee}, to extract linearly aligned points in a three-dimensional space. The algorithm will accelerate identifying tracklets from short video data, which is a fundamental routine in moving object detection.

The algorithm comprises two steps: creating a $k$-nearest neighbor ($k$-NN) graph and grouping elemental line segments (ELSs). The first part utilizes an efficient approximate $k$-NN graph construction method, NN-Descent \citep{dong_efficient_2011}. To compose the second part, we extend the fast line segment grouping method \citep{jang_fast_2002} to a three-dimensional space. The algorithm successfully identifies tracklets even when data are randomly decimated or are contaminated by distractors. We have confirmed that the algorithm works efficiently; the total elapsed time is approximately proportional to $N^{1.5}$, where $N$ is a representative size of the problem. There is, however, room for improvement for larger problems.

The algorithm is already implemented in the moving object detection system of the Tomo-e Gozen transient survey. Thanks to the high performance of \textit{tracee}, obtained data are processed almost in real-time. As of June 2021, 28 near-earth asteroids have been discovered. Although tracee is originally developed to detect asteroids, artificial satellites, and space debris from astronomical video data, the algorithm is potentially used in other fields, such as tracing ejecta's movement in an impact experiment.

The core functions of tracee are written in \texttt{C++}, and the interfaces for \texttt{Python} are implemented. \texttt{tracee} itself is implemented in \texttt{Python} and is made public as open-source software under the MIT license. The source code is hosted in a Bitbucket repository\footnote{\url{https://bitbucket.org/ryou_ohsawa/tracee/src/master/}}, and the package is available in the Python Package Index, \texttt{PyPI} \citep{ohsawa_tracee_2021}.

\section*{Acknowledgments}
\label{sec:acknowledgments}
This research has been partly supported by Japan Society for the Promotion of Science (JSPS) Grants-in-Aid for Scientific Research (KAKENHI) Grant Numbers 18K13599.

\appendix

\begin{table*}
\centering
\caption{A short example of \texttt{tracee}}
\label{tab:appendix:tutorial}
\begin{lstlisting}[language=Python]
#!/usr/bin/env python
# -*- coding: utf-8 -*-
import numpy as np
import tracee

# Load the data table.
# The table should have three columns.
filename = 'sample/mockdata_N006F030D500d00_b7e8.txt'
data = np.loadtxt(filename)

# Generate a parameter set.
param = tracee.default_parameters()
param.limit = 10  # Set the minimum number of points to 10.

# Call tracee.extract to identify tracklets.
# The function returns a list of Tracklet instances.
# The Tracklet instance contains
#   - tail: the starting point.
#   - head: the terminal point.
#   - members: a list of the associated points.
tracklet = tracee.extract(data, n_neighbor=3, param=param)
\end{lstlisting}
\end{table*}

\renewcommand{\thefootnote}{\relax}
\footnotetext{$^{*1}$Institute of Astronomy, Graduate School of Science, The University of Tokyo, 2-21-1 Osawa, Mitaka, Tokyo 181-0015, Japan \email{ohsawa@ioa.s.u-tokyo.ac.jp}}

\end{document}